\documentclass{iopart}

\usepackage[dvips]{graphicx,color}

\begin{document}
\title{Thermal-noise-limited underground interferometer CLIO}

\author{Kazuhiro Agatsuma$^1$, Koji Arai$^2$, Masa-Katsu Fujimoto$^2$,  
Seiji Kawamura$^2$, Kazuaki Kuroda$^1$, Osamu Miyakawa$^1$, Shinji Miyoki$^1$, 
Masatake Ohashi$^1$, Toshikazu Suzuki$^3$, Ryutaro Takahashi$^2$, Daisuke Tatsumi$^2$, Souichi Telada$^4$, 
Takashi Uchiyama$^1$, Kazuhiro Yamamoto$^5$, \\
and CLIO collaborators}%

\address{$^1$\ Institute for Cosmic Ray Research, University of Tokyo, Kashiwa, Chiba 277-8582, JAPAN}

\address{$^2$\ National Astronomical Observatory of Japan, Mitaka, Tokyo 181-8588, JAPAN}

\address{$^3$\ High Energy Accelerator Research Organization, Tsukuba, Ibaraki 305-0801, Japan}

\address{$^4$\ National Institute for Advanced Industrial Science and Technology, Tsukuba, Ibaraki 305-8563, Japan}

\address{$^5$\ Institut f\"{u}r Gravitationsphysik, Leibniz Universit\"{a}t
Hannover and Max-Planck-Institut f\"{u}r Gravitationsphysik
(Albert-Einstein-Institut),
Callinstrasse 38, D-30167 Hannover, Germany.}

\ead{agatsuma@icrr.u-tokyo.ac.jp}

\begin{abstract}
We report on the current status of CLIO (Cryogenic Laser Interferometer Observatory), 
which is a prototype interferometer for LCGT (Large Scale Cryogenic Gravitational-Wave Telescope). 
LCGT is a Japanese next-generation interferometric gravitational wave detector 
featuring the use of cryogenic mirrors and a quiet underground site. 
The main purpose of CLIO is to demonstrate a reduction of the mirror thermal noise by cooling the sapphire mirrors. 
CLIO is located in an underground site of the Kamioka mine, 1000 m deep from the mountain top, 
to verify its advantages. 
After a few years of commissioning work, we have achieved a thermal-noise-limited sensitivity at room temperature. 
One of the main results of noise hunting was the elimination of thermal noise 
caused by a conductive coil-holder coupled with a pendulum through magnets.
\end{abstract}

\submitto{\CQG}

\section{Introduction}
CLIO (Cryogenic Laser Interferometer Observatory)~\cite{Ohashi2003,Miyoki2004,Miyoki2006,Ohashi2008}, 
is a prototype for the next Japanese gravitational wave (GW) telescope project, 
LCGT (Large Scale Cryogenic Gravitational-Wave Telescope)~\cite{LCGT2006}, 
featuring the use of cryogenic mirrors and a quiet underground site. 
The main goal of CLIO is 
to demonstrate an improvement of sensitivity through a reduction of mirror thermal noise by cooling the sapphire mirrors.
The design sensitivity is limited by the mirror thermal noise and the suspension thermal noise around 100Hz, 
which will be reduced after cooling.
Through works on noise hunting, we achieved a thermal-noise-limited sensitivity at room temperature. 
One of main factors concerning the sensitivity improvement was to remove thermal noise 
due to a conductive coil-holder coupled with a pendulum through magnets.
Firstly, we report on the best displacement sensitivity we have achieved and noise hunting in recent works.
Next, we focus on verification of the thermal noise due to the coil-holder by theory and an experiment.

\section{Thermal-noise-limited sensitivity}

\subsection{Configuration of CLIO}
CLIO is located in Kamioka mine, which is 220 km from Tokyo, 
and lies 1000 m underground from the top of a mountain. 
LCGT is planed to be constructed in this area. 
This underground site is suitable for interferometric GW detectors,  
because the seismic noise is less than that in an urban area by about 2 orders~\cite{Araya1993,Yamamoto2006}. 
This merit is helpful to achieve the target sensitivity, and to obtain stability. 
Owing to this advantage in low frequency, observations for the Vela pulsar was performed in 2007~\cite{Akutsu2008}.

\begin{figure}[b]
  \begin{center}
\includegraphics[width=13cm,clip]{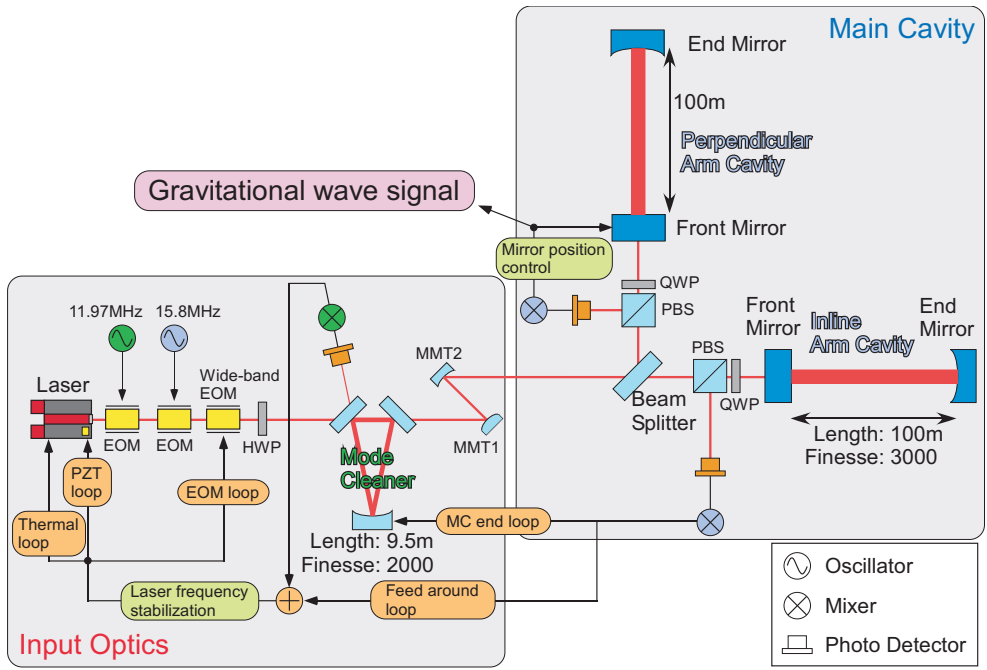}
  \end{center}
\caption{\label{fig:CLIO} Schematic view of CLIO;  
CLIO is a so-called locked Fabry-Perot interferometer, 
which has two 100 m Fabry-Perot cavities with a mode cleaner. 
The optics, like the lenses, the Faraday-isolators, and some wave plates are omitted in this figure.
Abbreviations denote: 
EOM, electro-optic modulator; 
HWP, half wave plate; 
QWP, quarter wave plate; 
MMT, mode matching telescope; 
and PBS, polarized beam splitter.
}
\end{figure}
Figure~\ref{fig:CLIO} shows a schematic view of CLIO. 
A laser beam (Innolight Inc. Mephisto) has a power of 2 watt and a 1064 nm wavelength. 
It is shaped into TEM00 through a mode cleaner (MC) cavity with a length of 9.5 m, 
and then injected to two 100 m length FP cavities after being divided by a beam splitter. 
These cavities are arranged in an L-shape. 
All of the returned (reflected) beam from each arm cavity is extracted
by an optical circulator formed by a $ \lambda /4 $ plate and a polarized beam splitter. 
Thus, this configuration does employ neither optical recombination by a Michelson interferometer nor optical recycling schemes.
The cavities are kept on resonance, called locked, by servo systems. 
The Pound-Drever-Hall technique~\cite{Drever1983} is employed as a readout method for displacement signals of the mirrors.
For that purpose, the phase modulations at 15.8 MHz and 11.97 MHz are used for the arm cavities and the mode cleaner cavity, respectively.
A multistage control system~\cite{Nagano2003} is applied for laser frequency stabilization, 
which has two cascaded loops of the MC and an arm cavity.
The inline arm is locked by controlling the frequency. 
The perpendicular arm is locked by controlling the mirror.
A differential displacement between two arm cavities, 
which corresponds to GW signals, can be obtained from a feedback signal to the coil-magnet actuators. 

The four mirrors of the two arm cavities are individually 
suspended by 6-stage pendulums, 
which include 4-stage blade springs and 2-stage wire suspensions 
for isolation from any seismic vibration.
The mirrors have weights of 1.8 kg (end) and 1.9 kg (front), made of sapphire. 
They are suspended by bolfur~\cite{Bolfur} wires at the last stage,
whose resonant frequency is 0.79 Hz. 
The pendulum has a resonant frequency of a primary mode of about 0.5 Hz. 
The angular alignments of all mirrors are tuned by movable stages on the suspensions.
Coil-magnet actuators are set for the front mirror in the perpendicular cavity only, 
so as to keep the optical path length of the cavity locked.

Until now, cryogenic systems have been developed~\cite{Tomaru2004,Tomaru2005,Li2005} and progressed~\cite{Uchiyama2006}.
Thus, cooling mirrors was realized~\cite{Yamamoto2007}. 
All sapphire mirrors were cooled down to the required temperature of under 20 K.
However, a thermal noise due to a conductive coil-holder, 
which appeared from 20 Hz to 300 Hz with a slope of $ f^{-2} $ ($ f $ was frequency) 
prevented the sensitivity from reaching the thermal noises of suspensions and mirrors, 
at both room temperature and the cool temperature~\cite{Yamamoto2007}. 
We could remove the extra thermal noise by noise hunting in recent works, and reached the design sensitivity at room temperature.

\subsection{Progress of noise hunting}
\begin{figure}[h]
  \begin{center}
\includegraphics[width=10cm,clip]{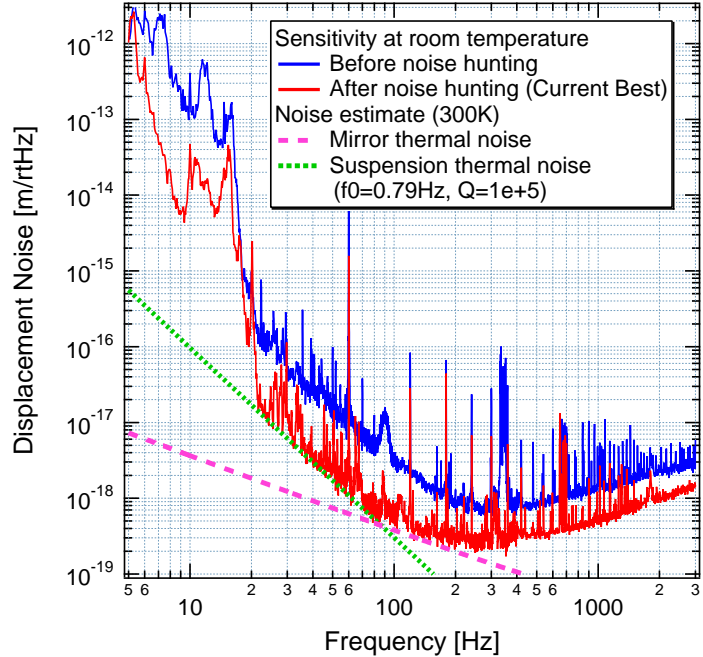}
  \end{center}
\caption{\label{fig:Best} Improvement of the sensitivity in 2008; 
Solid lines of red and blue are the measured sensitivity after noise hunting (Current best) and before noise hunting, respectively.
The green doted line is the estimated suspension thermal noise from wire material dissipations using Q-factors of the violin modes. 
The pink dash line indicates the calculated mirror thermal noise due to the substrate of the thermoelastic damping.
}
\end{figure}

CLIO displacement noise reached the predicted thermal-noise levels. 
Figure~\ref{fig:Best} shows the improvement of the displacement noise in 2008. 
the current best-floor sensitivity is $ 2.5 \times 10^{-19} $\,m/$ \sqrt{\rm{Hz}} $ at 250 Hz.
In the frequency region of 20 Hz to 80 Hz, 
the spectrum is close to the suspension thermal noise, 
which comes from wire-material dissipations of the structure damping (the internal damping~\cite{Saulson1990}). 
The theoretical prediction was calculated using the quality factor of a pendulum of $ 10^5 $, which was estimated from the measured violin Q.
The mirror thermal noise is also close to the sensitivity, 
and estimated from the bulk thermal noise of the thermoelastic damping. 
Details of this progress and estimations will be described in an independent article~\cite{Uchiyama2010}.

The noise hunting of CLIO interferometer has progressed at room temperature.
The improvement of the broad region from 20 Hz to 300 Hz came mainly from avoiding
a kind of pendulum thermal noise due to eddy currents in a coil-holder induced by magnets glued onto the mirror.
A beam centering adjustment also reduced noisy structures in this region.
By adjusting the beam centering, we could reduce the length fluctuation of the cavities coupled with a mirror-angle fluctuation.
In the high-frequency region, repairing the servo-circuits malfunction contributed to the reach of the shot noise. 
Using thinner suspension wires made of bolfur 
whose diameters were changed from 0.1 mm to 0.05 mm, 
shifted the violin modes to higher frequencies, 
so as to separate the skirts of violin modes from the mirror thermal noise region. 
One reason for the improvement at around 10 Hz was the removal of magnets on the upper mass for extended actuation, 
which had been shaken by damping magnets.

The most contributed thing for improving the sensitivity was to identify the pendulum thermal noise 
due to a coil-holder with a slope of $ f^{-2} $. 
Let us emphasize that this is not the suspension thermal noise.
The design sensitivity was calculated to be limited by the suspension thermal noise with a slope of $ f^{-5/2} $, 
which comes from wire-material dissipations of the internal damping~\cite{Saulson1990}.
On the other hand, 
the thermal noise due to a coil-holder was caused by 
mechanical losses by eddy-currents in the conductive coil-holder coupled with a pendulum.
The details are explained in the next section. 

\section{Thermal noise due to a coil-holder}

\subsection{Theoretical estimation}
Pendulum thermal fluctuation is caused by several kinds of dissipations coupled with the whole pendulum.
There are studies of those dissipations: 
an internal loss in the materials of suspension wires~\cite{Saulson1990,Gonzalez1995}, 
clamps of wires~\cite{Dawid1997}, 
residual air~\cite{Kajima1999},
coil-magnet actuators~\cite{Agatsuma2010_PRL},
and
a reference mass with coils~\cite{Cagnoli1998,Frasca1999}.
The pendulum thermal noise due to the reference mass with coils (the coil-holder in our case) is the most interesting in this section.
An oscillation of the pendulum gives rise to eddy currents by electromagnetic induction in the coil-holder. 
These currents generate Joule heat in the material of the coil-holder. 
According to the fluctuation-dissipation theorem (FDT)~\cite{Callen1951}, 
the dissipation of the Joule heat causes a thermal fluctuation of the pendulum through coupling with magnets glued onto the mirror. 

A quality factor of the pendulum, caused by the losses of conductive materials (reference masses) near the magnets, 
was derived~\cite{Cagnoli1998} as
\begin{equation}
  Q = \frac{m \omega _0}{2 \pi \sigma \left( \frac{3 \mu _0 \mathcal{M}}{4 \pi} \right)^2 J }  ,  \label{eq:Qh}
\end{equation}
which was verified by not only Cagnoli~\cite{Cagnoli1998}, but also Frasca~\cite{Frasca1999}.
Here, $ \omega_0 = 2 \pi f_0 $, $ f_0 $ is the pendulum frequency, 
$ m $ the mirror mass, $ \sigma$ the median conductivity of the materials, 
$ \mu_0 $ the permeability, $ \mathcal{M} $ the magnetic dipole moment of the magnet, 
and $ J $ a geometrical factor that depends on the shape of the conductor, given by
\begin{equation}
  J = \int_{z1}^{z2} \int_{r1}^{r2} \frac{r^3 z^2}{(r^2 + z^2)^{5}} drdz  .  \label{eq:J}
\end{equation}
Here, the conductor is assumed to have a center-holed cylindrical shape with an inner radius $ r1 $, 
outer radius $ r2 $, and length $ z2-z1 $ for simplicity.
The center of the magnet is placed at $ z=0 $, and the length from the magnet to the edge of a conductor is $ z1 $. 
The cause of thermal fluctuation is dissipation of the viscous damping, 
whose loss angle is $ \phi = \omega /( \omega _0 Q ) $, due to eddy currents in the conductor.
By using the FDT and applying the Q-factor of Eq.~(\ref{eq:Qh}), 
the thermal fluctuation of a pendulum (a harmonic oscillator) in a higher off-resonant region is approximately written as
\begin{equation}
  G= \frac{4 k_B T N}{m^2 \omega ^4} 2 \pi \sigma \left( \frac{3 \mu _0 \mathcal{M}}{4 \pi} \right)^2 J  .   \label{eq:Gh}
\end{equation}
$ \sqrt{G} $ is a one-sided power spectrum density, $ k_B $ the Boltzmann constant, 
and $ T $ the temperature of the conductors and the pendulum. 
$ N $ is the pair number of a magnet and a conductor that has a homogeneous shape around each magnet.

\begin{figure}[t]
  \begin{center}
\includegraphics[width=7.1cm,clip]{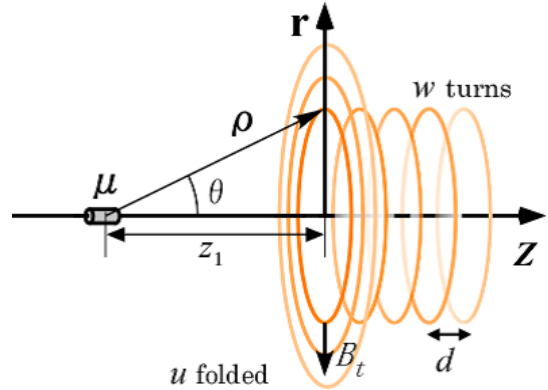}
  \end{center}
\caption{\label{fig:coordinate} Model of a coil-magnet actuator}
\end{figure}
In order to estimate the thermal noise due to the coil-holder by Eq.~(\ref{eq:Gh}) and the Q-factor of pendulum by Eq.~(\ref{eq:Qh}),
the magnetic dipole moment, $ \mathcal{M} $, is needed to know. 
However, the magnetic moment is not usually easy to measure. 
It is practicable to utilize a coil-magnet actuator coupling so as to estimate the magnetic moment.
By using one coil-magnet actuator, the force of $ F = \alpha I $ can be applied for the test mass.
The coupling factor, $ \alpha $, is the conversion efficiency between the current, $ I $, in a coil-circuit and the force, $ F $.
This coupling factor is related to the magnetic moment of a magnet by
\begin{equation}
  \alpha  = \frac{3\, \mu _0 \mathcal{M}}{ 2 } \sum_{s=0}^{u-1} \sum_{n=0}^{w-1} 
            \frac{(z_1 + d n)(r_1 + d s)^2}{((z_1 + d n)^2 + (r_1 + d s)^2)^{5/2}}  .   \label{eq:Mu}
\end{equation}
Here, the solenoidal coil, which consists of a conductive wire with a diameter of $ d $, 
is wound around a bobbin with a radius of $ r_{1} - d/2 $.
We approximately regarded the coil as a bunch of rings for modeling, as shown in Figure~\ref{fig:coordinate}.
On the heels of the first ring at a distance of $ z = z_1 $ from the magnet, 
rings of $ w $ turns are lined by a gap of $ d $ side by side ($ z = z_1 + d n ;\, n = 0,1,2,\cdots ,w-1 $).
When the coil is folded like layers to outside of its radius direction, 
it can be regarded as $ r = r_1 + d s ;\, s = 0,1,2,\cdots ,u-1 $ using a folding number of $ u $.
By measuring $ \alpha $, we can estimate the magnetic dipole moment, $ \mathcal{M} $, using Eq.~(\ref{eq:Mu}).
The coupling factor per one coil-magnet actuator, $ \alpha $, 
is yielded from the relation $ \alpha = A_{100} R_{\rm{c}} m (2 \pi \times 100)^2 / N_{\rm{c}} $.
$ A_{100} $ is the measured actuator response at 100 Hz, 
which is a transfer function from the driver input voltage to the mirror displacement. 
This response at the front mirror is measured using a Michelson interferometer constructed with front mirrors and the BS.
$ R_{\rm{c}} $ ($ R_{\rm{c}} = 50$ $ \Omega $ in CLIO) is the resistance of volt-current conversion in the coil-driver, 
and $ m $ is the weight of a test mass ($ m = 1.9 $ kg at the front mirror in CLIO). 
$ N_{\rm{c}} $ ($ N_{\rm{c}} = 4 $) is the pair number of a magnet and a coil.

\begin{figure}
  \begin{center}
\includegraphics[width=13.6cm,clip]{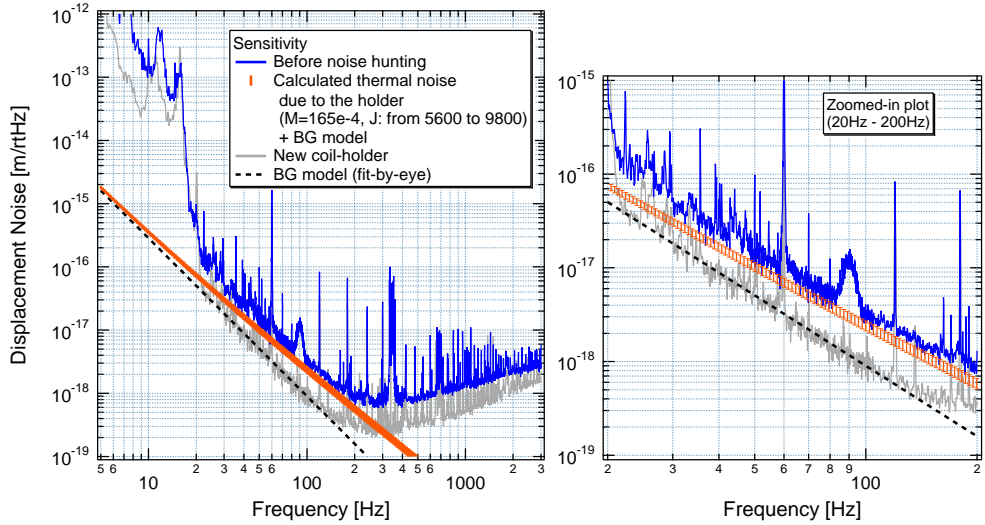}
  \end{center}
\caption{\label{fig:080515} 
Comparison between the measured spectrum and a theoretical calculation of the thermal noise due to the coil-holder; 
The solid lines of blue and gray indicates measured spectra 
with the previous coil-holder configuration and with the new coil-holder configuration, respectively.
The dash line denotes a background (BG) model as a part of sensitivity with the new coil-holder.
The orange band is the sum of the theoretical calculation of the thermal noise 
due to coil-holder material ($J$ of Eq.~\ref{eq:J}: from 5600 to 9800) and the BG model.
}
\end{figure}

The sensitivity of CLIO was improved by replacing a coil-holder. 
We tried to prove that the noise floor with the previous coil-holder came from the above thermal noise. 
However, the spectrum before noise hunting included some other noises. 
In order to take the other noises into account, 
an intermediate spectrum before the best sensitivity in Figure~\ref{fig:Best} was used as a background (BG) noise without eddy currents. 
Figure~\ref{fig:080515} shows a comparison between the measured spectrum before replacing the coil-holder 
and the calculation from Eq.~(\ref{eq:Gh}). 
The noise floor was almost matched with the calculation from 20 Hz to 200 Hz. 
The parameters for the estimates are indicated in Table~\ref{Table1}. 
The spectrum of the ``New coil-holder'' in Figure~\ref{fig:080515} is employed as the BG noise 
owing to the first measurement using the new coil-holder, 
which is close to the BG noise of the ``Before noise hunting''. 
For simplicity, we regarded a spectrum of the ``New coil-holder'' from 20 Hz to 100 Hz 
as the ``BG model'' with a slope of $ f^{-5/2} $ by fit-by-eye. 
The sensitivity in a high frequency region of Figure~\ref{fig:080515} was improved because the servo-circuit had been already mended. 
The spectrum of the ``New coil-holder'' include some changes not only the coil-holder but also other interferometer settings. 
Therefore a special experiment, which replaces only the coil-holder, is described in section 3.2.

\begin{table}
\caption{\label{Table1} Parameters; $ \alpha $ and $ \mathcal{M} $ are the average of all magnets. 
$ r2 $ at the previous coil-holder indicates the length from center of coil to a side or to a diagonal corner. 
The $ J $ values were calculated from $ r2 $ of two patterns (and the outer frame).}
\begin{indented}
\lineup
\item[]
  \begin{tabular}{@{}lll}
  \br
                       Parameter      & Previous coil-holder                          & New coil-holder                               \\ \mr
                Coil-holder (bobbin)  & Al                                            & Macor                                         \\ 
$ \sigma \,[\rm{\Omega ^{-1} m^{-1}}]$& \0\0\, $ 3.6 \times 10^7 $ & $ 10^{-13} $                                  \\ 
                       $ r1 $ [mm]    & \0\,  12                                      & ---                                           \\ 
                       $ r2 $ [mm]    & \0\,  15 -- 21                                & ---                                           \\ 
                       $ z1 $ [mm]    & \0\0\, 0                                      & ---                                           \\ 
                       $ z2 $ [mm]    & \0\,  20                                      & ---                                           \\ \mr
                        Coil          & Cu                                            & Cu                                            \\ 
                       $ w $ (turns)  & \0\,  22                                      & 15                                            \\ 
                       $ u $ (layers) & \0\0\, 1                                      & \, 2                                          \\ 
                       $ d $ [mm]     & \0\0\, 0.5                                    & \, 0.5                                        \\ 
                       $ r_1 $ [mm]   & \0\0\, 5.25                                   & \, 8.25                                       \\ 
                       $ z_1 $ [mm]   & \0\0\, 0                                      & \, 5                                          \\ \mr
                       Magnet         & Nd-Fe-B                                       &  Sm-Co                                        \\ 
                       Magnet size    & $ \phi $ 2mm $ \times $ 10mm                  & $ \phi $ 1mm $ \times $ 10mm                  \\ 
                      $ \alpha $ [N/A]& \0\0\, $ 3.6 \times 10^{-3} $                 & \, $ 4.8 \times 10^{-4} $                     \\ \mr
              Estimated value                                                                                                         \\ \mr
                 $ \mathcal{M} $ [J/T]& \0\0\, 0.0165                                 & \, 0.0034                                     \\ 
                $ J $ [1/$ \rm{m}^3 $]& 5600 -- 9000 (9800)                           & ---                                           \\
                  $ Q_{\rm{holder}} $ & \0\0\, $ 4.6 \times 10^{4} $                  & ---                                           \\ 
 $ \sqrt{G_{\rm{holder}}} $ at 100Hz [m/$ \sqrt{\rm{Hz}} $] & \0\0\, $ 2.5 \times 10^{-18} $ & ---                                    \\ 
  \br
  \end{tabular}
\end{indented}
\end{table}

The evaluation did not perfectly agree with the measured spectrum in Figure~\ref{fig:080515}. 
One reason is a limitation of the coil-holder modeling. 
The estimated geometrical factor, $ J $, is more uncertain than the other parameters, 
because the previous coil-holder had a cubic shaped structure around a coil 
in spite of the calculation limited by the cylindrical shape from Eq.~(\ref{eq:J}). 
A picture of this coil-holder is shown in Figure~\ref{fig:ExchangeHolder}. 
We estimated $ J $ to be a band line that has the lower limit of $ J=5600 $, and the upper limit of $ J=9800 $. 
$ J $ of 5600 is the lower limit and an underestimate 
because it does not include volume of the cubic corner beyond the cylindrical shape with $ r2 $ of 15 mm. 
$ J $ of 9000 is approximation of the whole cubic, 
which corresponds to $ r2 $ of a length from the center of the coil to a diagonal corner (21 mm). 
The upper limit is calculated as $ J=9800 $ by adding the effect of the outer frame (beyond the cubic around the coil) 
assuming a plate ($r1=21$ mm, $r2=32$ mm, $z1=10$ mm and $z2=20$ mm), 
which is rather an overestimate since the actual outer frame is not a plate shape. 
Numerical simulation is needed for more precise estimation using J of the coil-holder with a complex shape. 
Other reason is an error of the measured $ \alpha $, which is about 10 \%. 
The other reason is most likely a difference of BG noise in the region from 20 Hz to 50 Hz. 
In the ``Before noise hunting'', it is possible to inject the seismic noise via the coil-holder fixed on an optical stage 
or inject angular fluctuations of mirrors via an imperfection of a beam centering as mentioned in section 2.2.

\subsection{Experimental verification}
\begin{figure}[htb]
  \begin{center}
\includegraphics[width=7.1cm,clip]{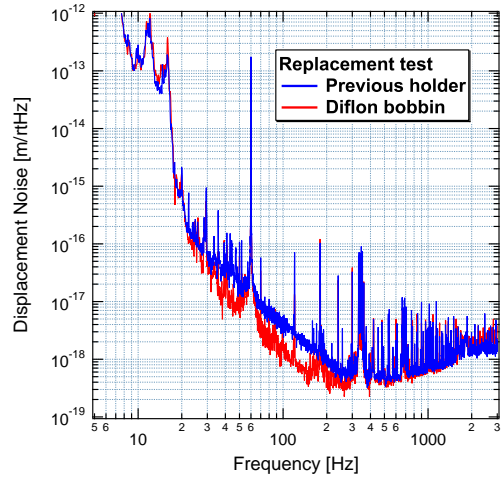}
  \end{center}
\caption{\label{fig:DiflonBobbin} Improvement of the sensitivity by replacing the coil-holder; 
The coil-holder made of aluminum was replaced with diflon bobbins, 
which were an electrical isolator. 
}
\end{figure}

We tested whether the sensitivity could be improved by replacing the coil-holder with an electrical isolator.
Figure~\ref{fig:DiflonBobbin} shows the result of that experiment.
The sensitivity was improved by replacing the coil-holder. 
The settings of the CLIO interferometer were not changed in the test, except for the coil-holder. 
That is why the noise floor with the previous coil-holder 
can be regarded as being the thermal noise due to the coil-holder coupled with the pendulum.
In this experiment, 
diflon bobbins (not the ``New coil-holder'' in Table~\ref{Table1} and Figure~\ref{fig:080515}) were employed so as to suppress eddy currents. 
The spectrum of the diflon bobbin was not employed as the BG noise in Figure~\ref{fig:080515} 
because the bobbin was supported by aluminum plates with a complex shape, which did not perfectly suppress eddy currents.
The aluminum was used as a material of the previous coil-holder for a reason of cryogenic compatibility as good thermal conduction. 
Further, a precise and quantitative identification of the pendulum thermal fluctuation due to viscous damping 
was performed using coil-magnet actuators~\cite{Agatsuma2010_PRL}. 

\newpage
\subsection{Solution}
\begin{figure}[hbt]
  \begin{center}
\includegraphics[width=13cm,clip]{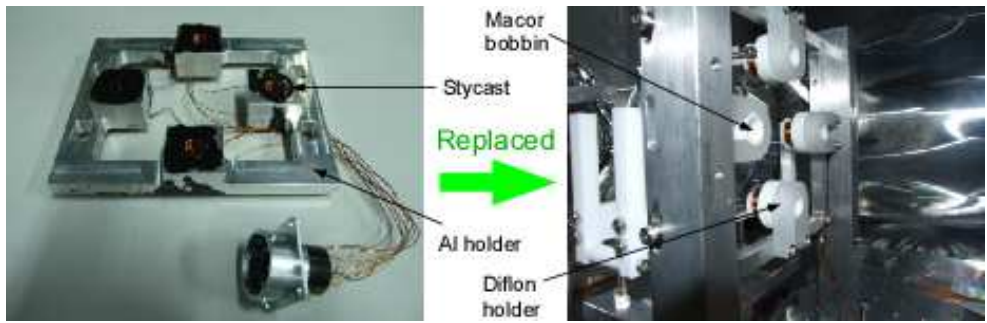}
  \end{center}
\caption{\label{fig:ExchangeHolder} Photographs of coil holders; 
Left: The previous coil-holder. 
Coils are surrounded by a coil-holder made of aluminum, and fixed into the coil-holder using stycast. 
Right: The new coil-holder.
Coils are wound on macor bobbins fixed on an aluminum frame through diflon holders.
}
\end{figure}

The coil-holder was redesigned so that the pendulum could sufficiently avoid mechanical losses due to eddy currents in the coil-holder.
The new coil-holder at room temperature is shown in Figure~\ref{fig:ExchangeHolder}. 
Coil bobbins are made of a macor (ceramic), which has an electrical conductivity of $ 10^{-13}$ $\rm{\Omega ^{-1} m^{-1}} $.
The aluminum frame is separated from the magnets apart.
The magnets were also changed from 2 mm to 1 mm of diameter, so that its magnetic moment could be reduced, 
and changed from Nd-Fe-B to Sm-Co of its material as a precaution against Barkhausen noise. 
The current best sensitivity at room temperature as shown in Figure~\ref{fig:Best} was accomplished in this coil-holder configuration.

For a cryogenic compatibility, 
the macor bobbins were replaced with aluminum nitride bobbins, which is an electrical isolator with a thermal conductivity.

\section{Summary}
CLIO is a prototype interferometer for LCGT, which is located in an underground site of the Kamioka mine.
After a few years of commissioning work, we achieved a thermal-noise-limited sensitivity at room temperature. 
The predicted suspension thermal noise and the mirror thermal noise were very close to the measured sensitivity.
The main factor concerning the sensitivity improvement was to remove thermal noise 
due to a conductive coil-holder coupled with a pendulum through magnets.
The experimental result was supported by the theoretical estimation.
This result became a direct verification in a noise spectrum of the estimation by Cagnoli and Frasca. 
We are ready to proceed to the cryogenic experiment, which is the main goal of thermal noise reduction in CLIO.
Currently, we are preparing to cool the mirrors as the next step.

\section{Acknowledgments}
This work was supported in part by
Global COE Program "the Physical Sciences Frontier", MEXT, Japan
and in part by a JSPS Grant-in-Aid for Scientific Research (No. 18204021).

\section{References}


\begin{thebibliography}{99}
\bibitem{Ohashi2003} Ohashi M \etal 2003 \emph{Class. Quantum Grav.} \textbf{20} S599
\bibitem{Miyoki2004} Miyoki S \etal 2004 \emph{Class. Quantum Grav.} \textbf{21} S1173
\bibitem{Miyoki2006} Miyoki S \etal 2006 \emph{Class. Quantum Grav.} \textbf{23} S231
\bibitem{Ohashi2008} Ohashi M \etal 2008 \emph{J. Phys.: Conf. Ser.} \textbf{120} 032008
\bibitem{LCGT2006} Kuroda K \etal 2006 \emph{Prog. Theor. Phys. Suppl.} \textbf{163} 54
\bibitem{Araya1993} Araya A, Kawabe K, Sato T, Mio N and Tsubono K 1993 \emph{Rev. Sci. Instrum.} \textbf{64} 1337
\bibitem{Yamamoto2006} Yamamoto K \etal 2006 \emph{J. Phys.: Conf. Ser.} \textbf{32} 418
\bibitem{Akutsu2008} Akutsu T \etal 2008 \emph{Class. Quantum Grav.} \textbf{25} 184013
\bibitem{Drever1983} Drever R W P, Hall J L, Kowalski F V, Hough J, Ford G M, Munley A J and Ward H 1983 \emph{Appl. Phys.} B \textbf{31} 97
\bibitem{Nagano2003} Nagano S \etal 2003 \emph{Rev. Sci. Instrum.} \textbf{74} 4176
\bibitem{Bolfur} A product of UNITICA http://www.unitika.co.jp/
\bibitem{Tomaru2004} Tomaru T, Suzuki T, Haruyama T, Shintomi T, Sato N, Yamamoto A, Ikushima Y, Koyama T and Li R 2004
\emph{Class. Quantum Grav.} \textbf{21} S1005
\bibitem{Tomaru2005} Tomaru T \etal 2005 Cryocoolers \textbf{13} 695
\bibitem{Li2005} Li R, Ikushima Y, Koyama T, Tomaru T, Suzuki T, Haruyama T, Shintomi T and Yamamoto A 2005 Cryocoolers \textbf{13} 703
\bibitem{Uchiyama2006} Uchiyama T \etal 2006 \emph{J. Phys.: Conf. Ser.} \textbf{32} 259
\bibitem{Yamamoto2007} Yamamoto K \etal 2008 \emph{J. Phys.: Conf. Ser.} \textbf{122} 012002
\bibitem{Saulson1990} Saulson P R 1990 \emph{Phys. Rev.} D \textbf{42} 2437
\bibitem{Uchiyama2010} Uchiyama T \etal manuscript in preparation.
\bibitem{Gonzalez1995} Gonz\'{a}lez G I and Saulson P R 1995 \emph{Phys. Lett.} A \textbf{201} 12
\bibitem{Dawid1997} Dawid D J and Kawamura S 1997 \emph{Rev. Sci. Instrum.} \textbf{68} 4600
\bibitem{Kajima1999} Kajima M, Kusumi N, Moriwaki S and Mio N 1999 \emph{Phys. Lett.} A \textbf{264} 251
\bibitem{Agatsuma2010_PRL} Agatsuma K \etal 2010 \emph{Phys. Rev. Lett.} \textbf{104} 040602
\bibitem{Cagnoli1998} Cagnoli G, Gammaitoni L, Kovalik J, Marchesoni F, and Punturo M 1998 \emph{Rev. Sci. Instrum.} \textbf{69} 2777
\bibitem{Frasca1999} Frasca S, Majorana E, Puppo P, Rapagnani P, and Ricci F 1999 \emph{Phys. Lett.} A \textbf{252} 11
\bibitem{Callen1951} Callen H B and Welton T A 1951 \emph{Phys. Rev.} \textbf{83} 34

\end{thebibliography}
\end{document}